\documentclass[12pt]{iopart}

\usepackage{iopams}
\usepackage{graphicx}

\begin{document}

\title[Ion pairs through long-range Rydberg molecules]{Formation of ultracold ion pairs through long-range Rydberg molecules}

\author{Michael Peper}
\address{Laboratory of Physical Chemistry, ETH Zürich, 8093 Zürich, Switzerland}
\author{Johannes Deiglmayr}
\address{Felix Bloch Institute, University of Leipzig, 04103 Leipzig, Germany}
\ead{johannes.deiglmayr@uni-leipzig.de}
\vspace{10pt}
\begin{indented}
\item[]September 2019
\end{indented}

\begin{abstract}
We propose a new approach to excite ion-pair states of ultracold dimers. The central idea is a two-step process where first long-range Rydberg molecules are formed by photoassociation, which are then driven by stimulated emission towards the ion-pair state, a process bearing features of a photo-induced harpooning reaction. We assess the feasibility of this approach through a detailed experimental and theoretical study on a specific system, $p$-wave-scattering dominated long-range Rydberg molecules in caesium, and discuss potential applications for the study of strongly correlated plasmas consisting of oppositely charged particles of equal or similar mass.
\end{abstract}

%
%
%
%
%

\section{Introduction}

In recent years, a solid understanding of the properties of long-range Rydberg molecules has been established and the developed theoretical tools explain experimental observations reliably, often even quantitatively \cite{shafferUltracoldRydbergMolecules2018,eilesTrilobitesButterfliesOther2019}. Long-range Rydberg molecules are bound states of an atom in a Rydberg state and a ground-state atom within the orbit of the Rydberg electron. The binding results from the Fermi-contact interaction \cite{fermiSopraSpostamentoPressione1934} of the (almost) free Rydberg electron scattering off the neutral ground-state atom. This interaction was first observed as a pressure-dependent shift of the transition frequencies in the spectroscopy of thermal vapours \cite{fuchtbauerUberIntensitatVerbreiterung1923,Amaldi1934}. In  Fermi's original pseudo-potential derivation, later extended by Omont~\cite{omontTheoryCollisionsAtoms1977} to include higher partial waves, the interaction potential between the Rydberg and the ground-state atom is given (in atomic units) by the expression
\begin{equation}\label{eq:FermiPwave}
  V = 2 \pi a_s \left|\Psi(R)\right|^2 + 6 \pi a_p \left|\nabla \Psi(R)\right|^2 \dots \;,
\end{equation}
where $\left|\Psi(R)\right|^2$ is the probability density of the Rydberg electron at the position $R$ of the ground-state atom, $\nabla \Psi(R)$ is the gradient of the wavefunction of the Rydberg electron, and the $s$-wave scattering length $a_s$ and the $p$-wave scattering volume $a_p$ are parameters directly derived from the electron-neutral interaction potential. The proportionality of the interaction strength to the probability density of the Rydberg electron leads to oscillatory interaction potentials supporting vibrational bound states, whose internuclear equilibrium separations are comparable to the size of the classical Rydberg orbit \cite{greeneCreationPolarNonpolar2000}.

We propose to exploit the properties of long-range Rydberg molecules for exploring a new route to highly-excited ion-pair states. Spectroscopy of ion-pair states and extrapolation of their Rydberg series to the ionization energy can yield precise values for electron affinities~\cite{herzbergRydbergSeriesIonization1972,beyerCommunicationHeavyRydbergStates2018}. Dissociation of weakly bound, ultracold ion pairs could provide a unique source of ultracold anions, \textit{e.g.}, for sympathetic cooling of antihydrogen~\cite{cerchiariUltracoldAnionsHighPrecision2018}. Ultracold clouds of dissociated ion pairs would also open up the ways to study neutral plasmas in an unexplored regime: in contrast to electron-cation plasmas studied previously~\cite{killianUltracoldNeutralPlasmas2007,lyonUltracoldNeutralPlasmas2016}, an anion-cation plasma, consisting of oppositely charged components with identical or similar masses, would feature dynamics on experimentally accessible time scales of microseconds~\cite{robicheauxSimulationsUltracoldNeutral2014a}.

The interaction potential between an anion and a cation at large distance is dominated by their Coulomb interaction just like in the case of a Rydberg atom where the negatively charged electron is bound to a positively charged ion core. The binding energy of vibrational levels in the ion-pair potential is given by the Rydberg formula
\begin{equation}\label{eq:hrEnergy}
    E = - h c \frac{R_{\mu}}{n_{\mathrm{HR}}^2} \;,
\end{equation}
where $R_{\mu}=R_{\infty}\frac{\mu}{m_{\mathrm{el}}}$ is the mass-corrected Rydberg constant, $n_{\mathrm{HR}}$ is an effective vibrational quantum number, $R_{\infty}$ is Rydberg's constant in wavenumbers, $\mu$ is the reduced mass of the two bound particles, $m_{\mathrm{el}}$ is the rest mass of the electron, and $h$ and $c$ are the Planck constant and the speed of light, respectively. For a bound cation-anion system, the value of $R_{\mu}$ is much larger than for the Rydberg states of atoms or molecules and the vibrational levels in the ion-pair-state potential have thus been named heavy-Rydberg states \cite{reinholdHeavyRydbergStates2005}. Because $R_{\mu}$ is very large, $n_{\mathrm{HR}}$ of a heavy-Rydberg state is much larger than the principal quantum number $n$ of an electronically-excited Rydberg state with the same binding energy.

At short internuclear distances, the electronic configuration of the ion-pair state mixes with excited electronic configurations of the neutral-parent dimer, possibly leading to strong perturbations of the electronic structure. Previous excitation schemes of molecular heavy-Rydberg states relied on such short-range mixings to excite the ion-pair states from the ground state of the parent molecule in a supersonic molecular beam (see, \textit{e.g.}, \cite{reinholdHeavyRydbergStates2005,molletDissociationDynamicsIonpair2010}) or in ultracold gases~\cite{kirranderApproachFormLongrange2013,marksonModelChargeTransfer2016}. However, these schemes have not yet yielded dense samples of heavy-Rydberg molecules, mostly because the exploited perturbations are weak and their accurate prediction is a significant challenge for current quantum-chemical methods.

\begin{figure}
\begin{center}
\includegraphics[width=0.95\linewidth]{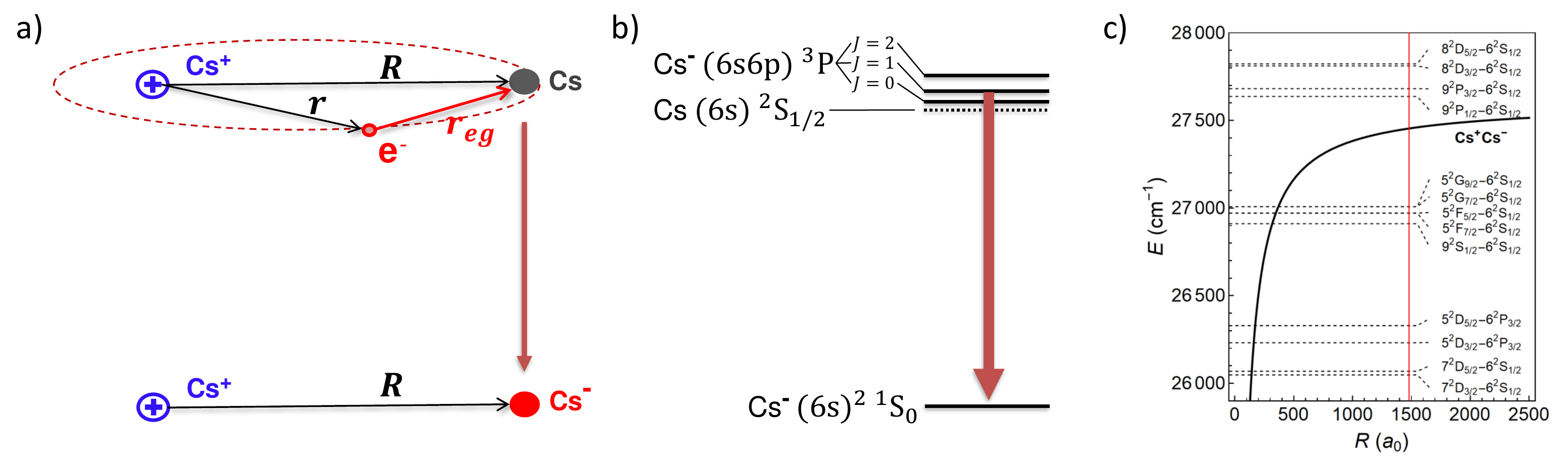}
\caption{\label{fig:Cs35p schematic idea} a) Schematic route for the excitation of heavy-Rydberg ion-pair states: a $p$-wave dominated scattering state of a Rydberg electron and a ground-state atom is deexcited to the electronic ground state of Cs$^-$ by stimulated emission driven via a mid-IR laser pulse; b) electronic structure of Cs$^-$: the dipole-allowed transition from the initial $^1$P$_1$ scattering state (a shape-resonance) to the $^1$S$_0$ electronic ground state is indicated by an arrow; c) Coulomb potential of the Cs$^-$\,-\,Cs$^+$ system and close-lying atomic asymptotes of Cs$_2$ \cite{NIST_ASD}, the red line marks the equilibrium separation $R \sim 1500\,a_0$ of a long-range Rydberg molecule in 35$^2$P$_{3/2}$.}
\end{center}
\end{figure}

Here we propose to excite the ion-pair states of a dimer via molecular long-range Rydberg states. In the past, we have studied extensively the long-range Rydberg states of Cs$_2$~\cite{sassmannshausenExperimentalCharacterizationSinglet2015,sassmannshausenLongrangeRydbergMolecules2016} and thus use them as model system here. In a semiclassical picture, the binding of long-range Rydberg molecules is based on the scattering of the almost-free Rydberg electron off a ground-state atom within its orbit. By stimulating the radiative transition from the electron-atom scattering state to the ground state of Cs$^-$, as depicted in Fig.~\ref{fig:Cs35p schematic idea}\,a), the long-range molecular Rydberg state is transferred into a bound state of Cs$^-$ and Cs$^+$. Note that in the case of a free electron, this transition would correspond to the reverse of the photo-detachment transition observed at a wavelength of $\sim 2.6\,\mu$m in Cs$^-$ (Fig.~\ref{fig:Cs35p schematic idea}\,b))~\cite{scheerExperimentalEvidenceThat1998}.

The selection rules for electric dipole transitions allow only for transitions between the $^1$P$_1$ $p$-wave component of the scattering state and the $^1$S$_0$ electronic ground state of Cs$^-$ (see Fig.~\ref{fig:Cs35p schematic idea}\,b)). Thus, the initial long-range Rydberg molecule has to be chosen such that its binding has a significant contribution from the $^1$P$_1$ scattering channel. The existence of bound states with this characteristic, called ``Butterfly molecules'' because of the shape of their spatial wavefunction, has been demonstrated both theoretically~\cite{hamiltonShaperesonanceinducedLongrangeMolecular2002} and experimentally~\cite{niederprumObservationPendularButterfly2016}. Transitions from a molecular Butterfly state to an ion-pair state may thus provide a novel route towards the excitation of heavy-Rydberg states~\cite{niederprumObservationPendularButterfly2016}.

The Franck-Condon principle will strongly favour transitions to Cs$^-$Cs$^+$ ion-pair states with an outer turning point close to the equilibrium distance of the molecular long-range Rydberg state (typically around 1000 to 2000\,$a_0$). The equilibrium distance of the ion-pair state is then about half of this distance, as known from the general properties of Rydberg states~\cite{gallagherRydbergAtoms1994}. The relevant part of the Coulomb-interaction potential is shown in Fig.~\ref{fig:Cs35p schematic idea}\,c) together with the close-lying atomic dissociation asymptotes of Cs$_2$. Inspection of Fig.~\ref{fig:Cs35p schematic idea}\,c) reveals a region of internuclear distances $R$ from $\sim 500\,a_0$ to $> 2500\,a_0$ where the potential-energy curve of the ion-pair state does not cross any dissociation threshold and should remain unperturbed.

In this article we present a detailed numerical study of this system, using Cs$_2$ as an example, emphasizing the importance of singlet-triplet mixing~\cite{sassmannshausenExperimentalCharacterizationSinglet2015} in the molecular long-range Rydberg state. We will separately discuss the following steps: \textit{i)} photoassociation of a pair of colliding atoms into metastable long-range Rydberg molecules with an electronic character that has significant contributions from $^1$P$_1$ scattering and \textit{ii)} driving stimulated emission from the long-range Rydberg to the ion-pair state. This photoinduced charge-transfer process bears features of a harpoon reaction, where the Rydberg atom captures the ground-state atom by donating its electron to the neutral and a more strongly bound ion pair is formed. Note that the direct photoassociation into an electronic state with ion-pair character is only possible at much shorter internuclear distances than considered here, where covalent and ionic molecular states are strongly mixed~\cite{kirranderApproachFormLongrange2013,marksonModelChargeTransfer2016}.

\section{Photoassociation of butterfly-like long-range Rydberg molecules}

In 2002, Hamilton \textit{et al.} predicted that shape resonances in $p$-wave electron-alkali-atom scattering cause molecular long-range Rydberg states with large binding energies and electronic wavefunctions resembling (in a specific graphical representation) the shape of a butterfly~\cite{hamiltonShaperesonanceinducedLongrangeMolecular2002}. Shape resonances result from metastable scattering complexes bound behind a centrifugal barrier which can strongly modify the scattering interactions of colliding particles. On resonance, the scattering electron is temporarily bound in an excited state of the anion which provides an opportunity to deexcite the collision complex to the anionic ground state. Experimental signatures of butterfly molecules have been observed in rubidium close to the 25\,$^2$P$_{3/2}$ asymptote, where an analysis of the rotationally-resolved photoassociation spectra revealed that the excited molecular Rydberg states had internuclear separations between 100 and 300\,$a_0$~\cite{niederprumObservationPendularButterfly2016}. More recently, molecular bound states with strong contributions to their binding from $p$-wave scattering have been observed close to $n$\,$^2$S$_{1/2}$ \cite{engelPrecisionSpectroscopyNegativeIon2019} and $n$\,$^2$D$_{j}$ \cite{maclennanDeeplyBound24D2019} Rydberg states of rubidium. The $p$-wave character of these ``butterfly-like'' states, bound just below low-$l$ Rydberg states, is smaller than the one of the true butterfly molecules discussed above. However, the equilibrium distances of these molecules are on the order of 1000\,$a_0$, which makes such states a more promising intermediate state for transferring an ultracold atomic gas into a gas of weakly bound ion pairs.

\begin{figure}
\begin{center}
\includegraphics[width=0.48\linewidth]{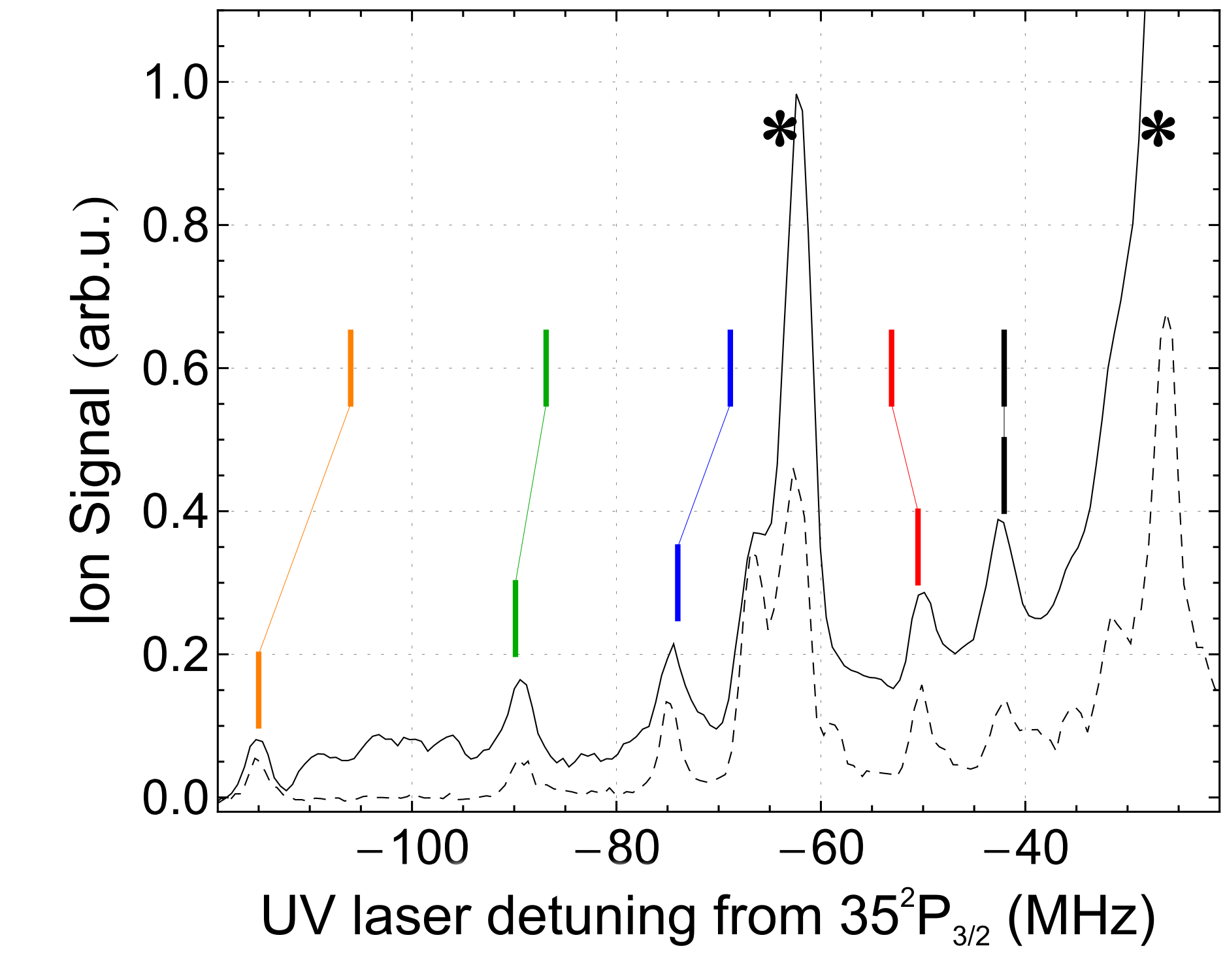}
\includegraphics[width=0.48\linewidth]{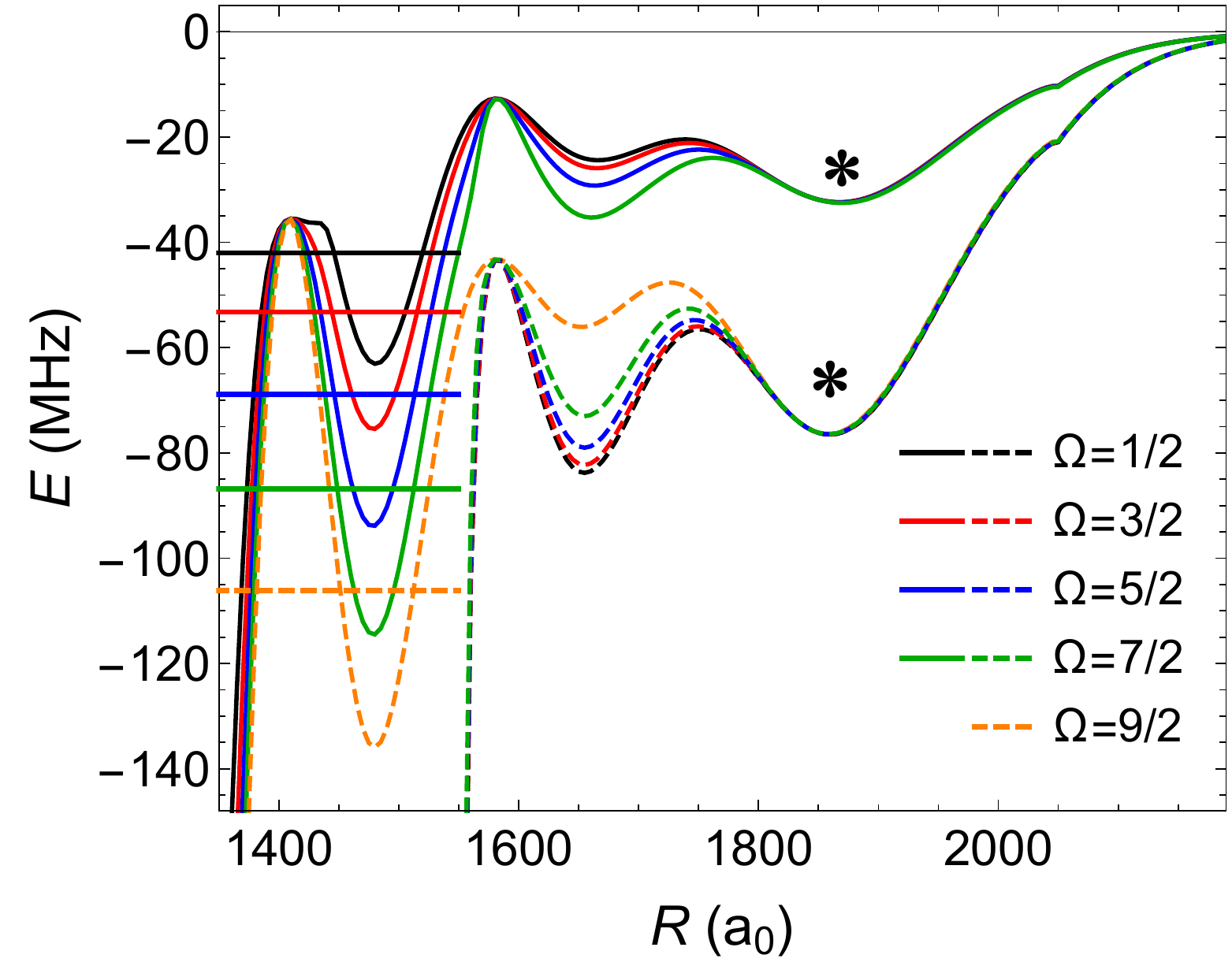}
\caption{\label{fig:Cs35p} Left panel: experimental photoassociation spectra in the vicinity of the 6$^2$S$_{1/2}$($F=4$)-$n^2$P$_{3/2}$ asymptote, comparing the yield of Cs$^+$ ions resulting from field-ionization of Rydberg atoms and molecules (solid line) to the yield of Cs$_2^+$ ions resulting from associative ionization of Rydberg molecules (dashed line). Right panel: calculated potential energy curves with $^{1,3}\Sigma^+$ symmetry (solid lines) and $^{3}\Sigma^+$ symmetry (dashed lines). Asterisks mark experimental resonances (left panel) arising from photoassociation of molecules localized in the outer-most wells (right panel). Resonances attributed to photoassociation of $p$-wave dominated molecules are marked in the left panel by coloured lines, where the lower vertical line indicates the experimental line position and the upper vertical line the calculated binding energy of the assigned molecular state.  The calculated binding energies are also indicated in the right panel by horizontal lines.}
\end{center}
\end{figure}

Here we investigate $p$-wave-scattering dominated trilobite-like molecular states in caesium close to $n$\,$^2$P$_{3/2}$ states. The electron-caesium scattering features broad $p$-wave shape resonances close to zero collision energy~\cite{scheerExperimentalEvidenceThat1998}, and thus strong contributions from $p$-wave scattering to the binding of long-range Rydberg molecules are to be expected. The experimental setup has been described previously~\cite{sassmannshausenExperimentalCharacterizationSinglet2015}: we release ultracold samples of Cs atoms from a compressed magneto-optical trap (MOT) with typical densities of 10$^{12}$ atoms/cm$^3$ and temperatures of 40 $\mu$K, as determined by absorption imaging. The atoms are then prepared in a specific hyperfine state ($F=3$ or 4) of the 6\,$^2$S$_{1/2}$ ground state by optical pumping. After turning off all optical and magnetic trapping fields, single-photon photoassociation into long-range Rydberg molecules is performed. The UV radiation driving the photoassociation transitions is obtained by frequency-doubling the output of a continuous-wave (cw) single-mode ring dye laser at $\sim$639 nm and forming pulses of 3\,$\mu$s length using an acousto-optic modulator. The laser frequency is calibrated with the help of a wavemeter and a frequency comb~\cite{deiglmayrPrecisionMeasurementIonization2016}. The formation of Rydberg molecules is detected by applying pulsed electric potentials to a pair of electrodes surrounding the photoexcitation region, creating an electric field with sufficient magnitude to field ionize the Rydberg molecules. Field-ionization of long-range Cs$_2$ Rydberg molecules produces an electron, a Cs$^+$ ion, and a neutral Cs atom, because the molecule instantaneously dissociates upon removal of the Rydberg electron, which mediates the binding interaction. We detect the resulting Cs$^+$ ions by accelerating them towards a micro-channel plate (MCP) detector and sample the detector current using a fast oscilloscope. The extraction field also accelerates Cs$_2^+$ ions, which result from vibronic autoionization of long-range Rydberg molecules~\cite{niederprumGiantCrossSection2015}, towards the detector. Cs$^+$ and Cs$_2^+$ ions are easily separated by their different arrival times at the detector. The respective ion-yields, plotted in Figure~\ref{fig:Cs35p}\,a), are obtained by integrating the oscilloscope traces in narrow integration windows around the mean arrival times of Cs$^+$ and Cs$_2^+$ ions.

In Fig.~\ref{fig:Cs35p} (left panel), a photoassociation spectrum taken in the vicinity of the 35\,$^2$P$_{3/2}$ asymptote is presented. The right-hand panel of this figure depicts the molecular structure close to this asymptote, calculated on the basis of the Hamiltonian of Eiles and Greene~\cite{eilesHamiltonianInclusionSpin2017}. This Hamiltonian includes $s$- and $p$-wave scattering as well as all relevant interactions of electronic and nuclear spins. The electron-Cs scattering parameters are adopted from Ref.~\cite{marksonTheoryUltralongRangeRydberg2016}, and the atomic basis set includes the manifolds $n=31-33$ and all low-$\ell$ states within this range. The inclusion of the nuclear spin of the ground-state atom leads to the appearance of two classes of molecular states: $^3\Sigma^+$ states with a binding resulting only from scattering in the triplet states of Cs$^-$, and $^{1,3}\Sigma^+$ states with contributions from both triplet and singlet scattering~\cite{andersonAngularmomentumCouplingsLongrange2014,sassmannshausenExperimentalCharacterizationSinglet2015}. These labels are only approximate because, \textit{e.g.}, the spin-orbit interaction mixes $\Sigma$ and $\Pi$ character. The remaining good quantum number in the non-rotating molecule is $\Omega$, the projection of the total angular momentum (excluding rotation) on the internuclear axis.

Two strong resonances marked by asterisks are assigned to the lowest vibrational state in the outermost well of the $^3\Sigma^+$ and $^{1,3}\Sigma^+$ states, which are almost degenerate in $\Omega$. The resonance positions agree very well with our previous analysis of these molecular resonances over a large range of $n$ values considering only $s$-wave scattering~\cite{sassmannshausenExperimentalCharacterizationSinglet2015}. The calculation reveals that the degeneracy of the potential-energy curves in $\Omega$ is lifted at shorter internuclear separations, where contributions from $p$-wave scattering become important. The depth of the inner wells of the $^{1,3}\Sigma^+$ states centered around 1480~$a_0$ becomes strongly $\Omega$-dependent. These inner wells support only a single bound state each, whose zero-point energy is determined from a harmonic approximation of the potential. The bound states in the $\Omega$-split potentials are expected to yield a regular series of resonances. Indeed, a regular series of molecular resonances is observable in the experimental photoassociation spectra close to 35\,$^2$P$_{3/2}$ (Fig.~\ref{fig:Cs35p}, left panel). The calculated series of binding energies agrees qualitatively with the observed structure of photoassociation resonances. The tentative assignment shown in Fig.~\ref{fig:Cs35p} reduces the overall deviation, but a more detailed theoretical study is necessary to explain the remaining deviations, as well as the mechanism for the observed autoionization limiting the molecular lifetimes to few microseconds. Very similar electronic structures have been predicted below the 32$^2$P$_{3/2}$ \cite{marksonTheoryUltralongRangeRydberg2016} and the 33$^2$P$_{1/2}$ \cite{eilesHamiltonianInclusionSpin2017} asymptote in caesium, which we have confirmed in our calculations. The choice of $n$ determines the range of binding energies and equilibrium distances of the addressed ion-pair states, which scale as $1/n^2$ and $n^2$, respectively. Simple estimates based on the normalisation of the Rydberg wavefunction, the $n^2$ dependence of the classical Rydberg orbit, and the pair-distance distribution in a homogeneous gas,k show that photoassociation rates should increase linearly with $n$ and the rates for stimulated emission into the ion-pair state, discussed in the next section, should scale as $n^{-3}$. We chose to continue our exploration using molecules bound close to the 35\,$^2$P$_{3/2}$ asymptote as an example from which rough estimates for other values of $n$ can be obtained.

\section{Transfer to an ion-pair state}

\begin{figure}
\begin{center}
\includegraphics[width=0.6\linewidth]{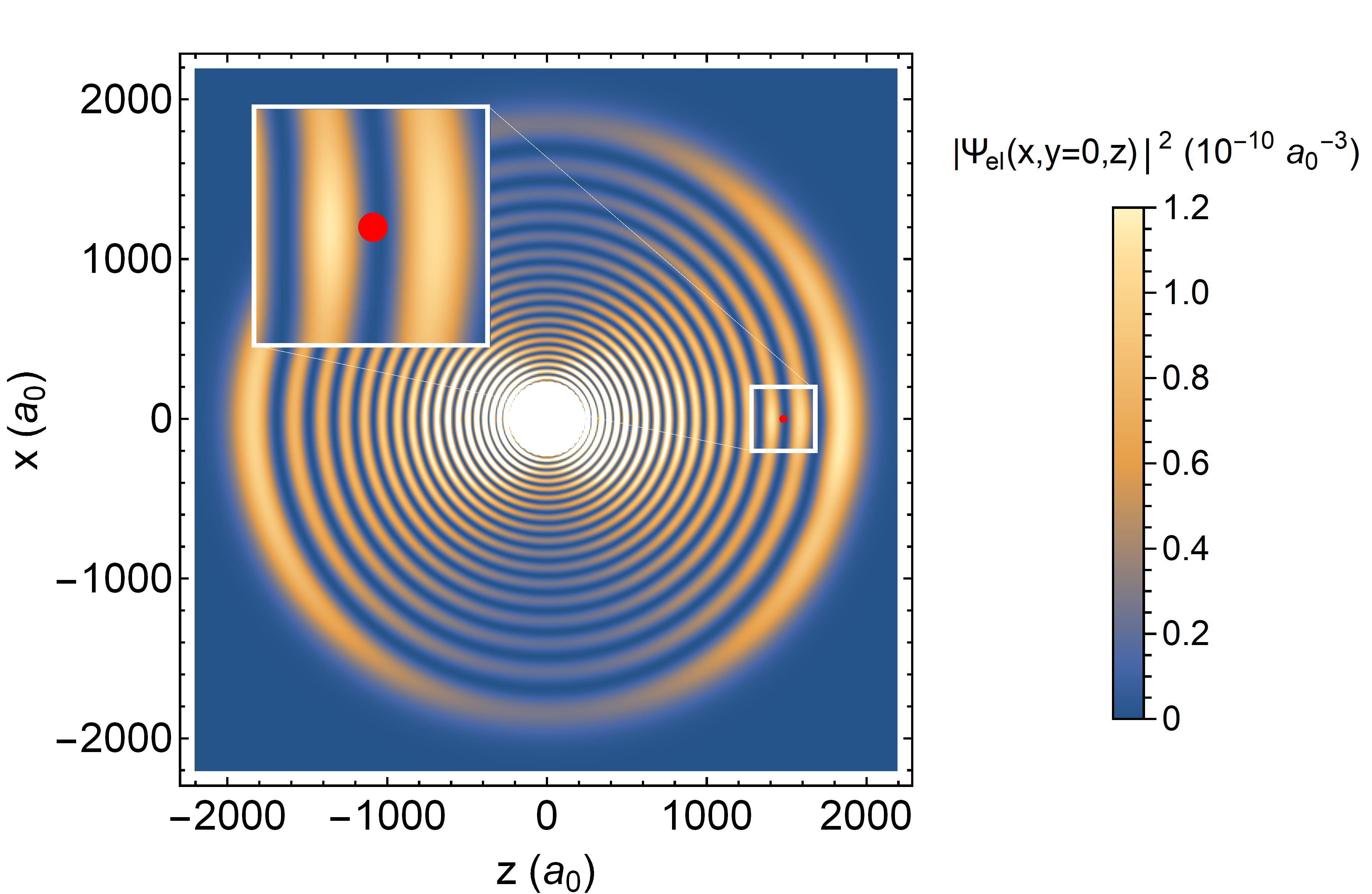}
\caption{\label{fig:Cs35pElEfct} Cut through the electronic-density distribution of the $^{1,3}\Sigma^+$ state with $\Omega=5/2$ (solid blue line in the right panel of Fig.~\ref{fig:Cs35p}) at $R = 1480\,a_0$ in a plane containing the internuclear axis (chosen along the $z$-axis). A red dot marks the position of the perturbing ground-state atom. The inset shows the region around the perturber at an enlarged scale. }
\end{center}
\end{figure}

Encouraged by the qualitative agreement between the theoretical model and the experimental observations presented above, we use the calculated molecular structure to evaluate the feasibility of transferring the long-range Rydberg molecules into an ion-pair state. An analysis of the calculated electronic wavefunctions around $R \sim 1480\,a_0$ shows that these states have strong contributions both from singlet- and triplet-scattering channels, and that the interaction is indeed dominated by $p$-wave scattering. An investigation of the spatial component of the electronic wavefunctions at the molecular equilibrium separation, depicted for $\Omega=5/2$ in Fig.~\ref{fig:Cs35pElEfct}, reveals that these bound states are localized close to the position of maximal gradient of the wavefunction of the Rydberg electron, \textit{i.e.}, at a node of its probability density, as expected for a $p$-wave dominated molecular state, see Eq.~(\ref{eq:FermiPwave}). While this property facilitates the optical dipole transitions from the molecular long-range Rydberg state to ion-pair states, it also results in small transition dipole moments.

\begin{figure}[h]
\begin{center}
\includegraphics[width=0.8\linewidth]{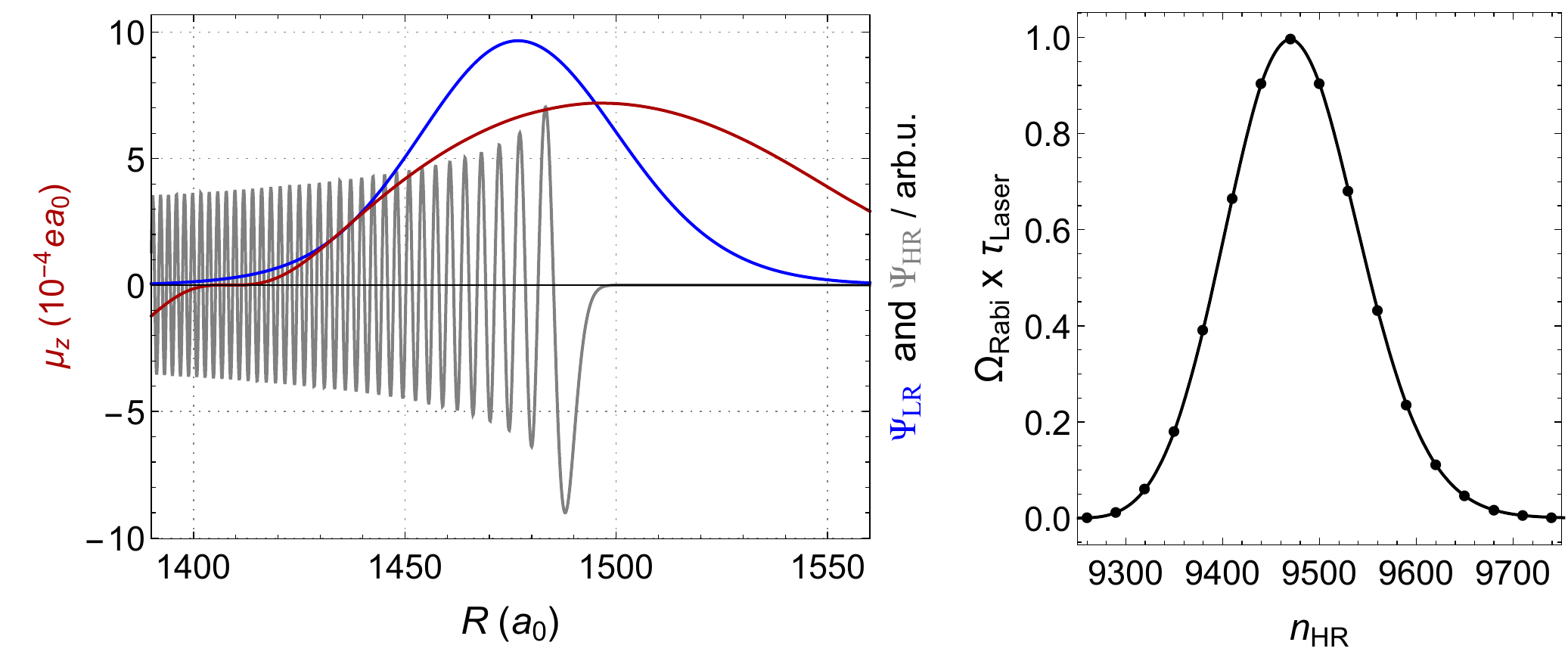}
\caption{\label{fig:Cs35pDump} Left panel: (red line, left axis) calculated $R$-dependent transition dipole moment ($\Delta m_l = 0$) between the electronic state $^{1,3}\Sigma^+$ ($\Omega=5/2$), whose spatial wavefunction is presented in Fig.~\ref{fig:Cs35pElEfct}, and the $^1$S$_0$ ground state of Cs$^-$; (blue line, right axis) vibrational wavefunction of the long-range Rydberg molecule in $v=0$; (gray line, right axis) vibrational wavefunction of a Cs$_2$ heavy-Rydberg state in $n_{\mathrm{HR}}= 9\,500$. Right panel: Calculated product of Rabi rate and pulse width ($\tau_{\mathrm{laser}}=5$\,ns) as a function of the vibrational level of the final heavy-Rydberg state $n_{\mathrm{HR}}$ for a pulse energy of 10\,$\mu$J and a beam waist radius of 300\,$\mu$m.}
\end{center}
\end{figure}

In order to estimate the rates for stimulated emission, we focus on these $(\Omega)\,^{1,3}\Sigma^+\,(v=0)$ states (vibrational wavefunction for $\Omega=5/2$ shown in Fig.~\ref{fig:Cs35pDump}). The $R$-dependent transition dipole moment between the long-range Rydberg state and the ion-pair state is obtained by numerical evaluation of the matrix element of the dipole operator between the wavefunction of the Rydberg electron (see, \textit{e.g.}, Fig.~\ref{fig:Cs35pElEfct}) and the $(6s)$-single-electron wavefunction of Cs$^-$ obtained from a Hartree-Fock-based calculation~\cite{cowanTheoryAtomicStructure1981} in cylindrical coordinates centered at the position of the ground-state atom under conservation of nuclear and electronic spins. Note that in the employed electric dipole approximation, only the singlet component of the long-range Rydberg molecules wavefunction couples to the $^1$S$_0$ ground state of Cs$^-$. This neglects the mixing of $^3$S$_1$ and $^1$P$_1$ scattering channels by the spin-orbit interaction in Cs$^-$~(\cite{khuskivadzeAdiabaticEnergyLevels2002}) and the calculation is thus likely to underestimate the transition rate. Only parallel transitions ($\Delta m_l = 0$) yield finite transition-dipole moments, perpendicular transition-dipole moments ($|\Delta m_l| = 1$) are strongly suppressed. This is not surprising, because in the ground state, $m_l$ is zero and the $|m_l| = 1$ component of the Rydberg-electron's wavefunction has a node on the internuclear axis which was chosen as the $z$-axis. The dependence of the transition dipole moment on $\Omega$ is found to be weak, with $\mu_z$ decreasing by about a factor of two from $\Omega=1/2$ to $\Omega=7/2$. We thus chose the intermediate $\Omega=5/2$ as exemplary state to continue our analysis.

Vibrational wavefunctions of initial and final molecular states as well as the expectation value of the transition-dipole moment are derived by numerical integration using Numerov's algorithm. The predicted spontaneous decay rate (Einstein $A$ coefficient $\ll{1\,\textrm{s}^{-1}}$) is much smaller than the rate for radiative decay of the long-range Rydberg molecules, which can be well approximated by the decay rate for an isolated Rydberg atom (\textit{e.g.}, for Cs 35$^2$P$_{3/2}$, Ref.~\cite{ovsiannikovRatesBlackbodyRadiationinduced2011} calculated a total decay rate, including spontaneous decay and transitions induced by black-body radiation, of 28\,kHz). It is therefore necessary to enhance the transition rate, \textit{e.g.}, by stimulated emission in an external light field. The required wavelength lies in the mid-infrared region of the optical spectrum (around 2.6~$\mu$m) where intense, broadly tunable cw laser sources are rare. We thus evaluate the transition rates for pulsed laser sources, such as a pulse-amplified cw laser system combined with difference-frequency mixing~\cite{jacovellaInfraredSpectroscopyMolecular2016}, and calculate the Rabi frequency for the driven transition. For realistic laser parameters given in the caption of Fig.~\ref{fig:Cs35pDump}, the product of Rabi frequency and pulse length reaches unity, indicating that the transition can be saturated.

Within the Franck-Condon window, final heavy-Rydberg states between $n_{\mathrm{HR}}\sim 9\,350$ and $n_{\mathrm{HR}}\sim 9\,600$ can be addressed, corresponding to an energy span of about 240\,GHz. The expected vibrational spacing of the addressed heavy-Rydberg states is around 1~Ghz, which is well resolvable by a mid-IR laser pulse of a few nanoseconds length (transform-limited bandwidth typically below 150~MHz). Transitions will be observable by a loss in the yield of field-ionized Rydberg molecules and the appearance of Cs$^-$ anions on resonance. The addressed heavy-Rydberg states of Cs with $n_{\mathrm{HR}}\sim 10\,000$ have a significantly increased sensitivity to electric fields: classically, the cation-anion system at an internuclear separation of $R\sim 1\,500 a_0$ constitutes an electric dipole with a dipole moment $\mu=1\,500 e a_0$, where $e$ is the elementary charge. Assuming random orientations of this dipole in an external electric field of 10\,mV/cm, the resulting transition broadening amounts to $\sim$40\,MHz. The Inglis-Teller limit, \textit{i.e.}, the electric-field strength where neighbouring Rydberg manifolds start to overlap, is reached at a field strength of $\sim$250\,mV/cm~\cite{reinholdHeavyRydbergStates2005}. Given typical residual fields in current experiments of $\leq10$\,mV/cm~\cite{peperPrecisionMeasurementIonization2019} and the envisioned laser linewidth of $\sim$150\,MHz, the electric-field induced broadening should not hinder the spectroscopic study of heavy-Rydberg states.

Direct photoionization of the Rydberg molecule by the mid-IR radiation competes with stimulated emission. The photoionization rates for a long-range Rydberg molecule are very similar to the rates for isolated Rydberg atoms, which have been calculated by Ovsiannikov \textit{et al.}~\cite{ovsiannikovRadiativeLifetimePhotoionization2011}. Based on their calculations, the ionization probability for the laser parameters given in the caption of Fig.~\ref{fig:Cs35pDump} (a Gaussian pulse with temporal full width at half maximum of 5\,ns and a peak intensity of about $2\cdot10^6$\,W/cm$^2$) is about 3\%, significantly smaller than the probability for stimulated decay (which is about unity for the given parameters).

The lifetime of heavy-Rydberg states is in general limited by couplings to valence states of the neutral parent species at short internuclear distances, which can cause dissociation or autoionization of the ion-pair state as observed in other heavy-Rydberg systems \cite{reinholdHeavyRydbergStates2005,molletDissociationDynamicsIonpair2010}. Estimating lifetimes of the excited ion-pair states goes beyond the scope of this paper. However, the lifetimes should increase with increasing rotational angular momentum of the ion pair.

\section{Discussion and outlook}

In this article we have presented a novel route to excite heavy-Rydberg states with high values of $n_{\mathrm{HR}}$ via photoassociation of long-range Rydberg molecules and stimulated emission into the ion-pair state. By performing detailed calculations for a specific system (molecular states close to the 6\,$^2$S$_{1/2}$-35\,$^2$P$_{3/2}$ asymptote in caesium), we find this approach to be feasible for realistic experimental parameters. We note that there is nothing specific about the chosen system, except for the criteria discussed above, and that we are confident that the same approach can be implemented in a variety of systems, employing photoassociation close to, \textit{e.g.}, $n$\,$^2$S$_{1/2}$, $n$\,$^2$P$_{j}$ or $n$\,$^2$D$_{j}$ asymptotes in homonuclear (Cs$_2$, Rb$_2$, K$_2$, \dots) or heteronuclear alkali dimers~\cite{eilesFormationLongrangeRydberg2018}, as long as the formed long-range Rydberg states have significant contributions from $p$-wave scattering. The scheme can be similarly applied to more complex chemical species, such as, \textit{e.g.}, dimers composed of the alkaline earth metals with stable anions (Ca, Sr, Ba). For example, the formation of long-range Sr$_2$ Rydberg molecules has already been demonstrated~\cite{desalvoUltralongrangeRydbergMolecules2015}. Because the ground state of alkaline earth anions is a $^2$P$_j$ state, efficient transfer from $s$-wave-scattering dominated long-range Rydberg molecules to ion-pair states can be expected in these systems. The lower electron affinity of alkaline-earth atoms will, however, require to stimulate the transitions by radiation in the terahertz region.

The approach presented in this work opens up the ways to study structure and dynamics of heavy-Rydberg states in a previously unexplored region of internuclear separations. As discussed in the introduction, a dense gas of ion pairs may transform into a plasma with equal-mass charges (for homonuclear systems) or charges with variable mass ratios (for heteronuclear systems). This transformation may occur spontaneously, as routinely observed in dense gases of Rydberg atoms \cite{tannerManyBodyIonizationFrozen2008,robert-de-saint-vincentSpontaneousAvalancheIonization2013, sassmannshausenPulsedExcitationRydbergatompair2015,forestExpansionUltracoldRydberg2018}, or be stimulated by DC-electric-field-induced dissociation \cite{molletDissociationDynamicsIonpair2010} or multi-photon dissociations by low-frequency radio-frequency (RF) fields \cite{arakelyanIonizationNaRydberg2016}. A theoretical study by Robicheaux~\textit{et.~al}~\cite{robicheauxSimulationsUltracoldNeutral2014a} predicts the equilibration dynamics of plasmas with equal-mass charges to be on experimentally accessible timescales in the range of microseconds. The non-equilibrium dynamics of such a strongly coupled system are expected to exhibit strong spatial correlations between particles and show physical phenomena relevant to the description of giant gas planets like Jupiter or fusion devices \cite{killianUltracoldNeutralPlasmas2007}. The neutral plasma might even be confined in the RF field of a hybrid atom-ion trap~\cite{deiglmayrReactiveCollisionsTrapped2012}, because the RF parameters can be easily chosen such that the masses of all components fall within the trap's stability window~\cite{westerRadiofrequencyMultipoleTraps2009}.

\ack
We thank Fr\'ed\'eric Merkt for his continuous and generous support. We thank Matthew Eiles, Chris Greene, Francis Robicheaux, Christian Fey, Frederic Hummel, Herwig Ott, and Hossein Sadeghpour for fruitful discussions. This work is supported financially by the ETH Research Grant ETH-22 15-1 and the NCCR QSIT of the Swiss National Science Foundation.\\


\end{document}